\documentclass[12pt]{JHEP}
\usepackage{graphics}
\usepackage{cite}
\newcommand{\beq}{\begin{equation}}
\newcommand{\eeq}{\end{equation}}
\newcommand{\bea}{\begin{eqnarray}}
\newcommand{\eea}{\end{eqnarray}}

\renewcommand{\d}{\delta}
\renewcommand{\l}{\lambda}

\renewcommand{\b}{\beta}
\renewcommand{\a}{\alpha}
\renewcommand{\k}{\kappa}
\renewcommand{\ni}{\noindent}
\newcommand{\n}{\nu}
\newcommand{\m}{\mu}
\renewcommand{\r}{\rho}

\newcommand{\s}{\sigma}
\newcommand{\A}{{\cal A}}

\newcommand{\U}{{\cal U}}

\newcommand{\tr}{\widetilde{\rho}}

\newcommand{\e}{\epsilon}

\newcommand{\oh}{\frac{1}{2}}

\newcommand{\dg}{\dagger}
\newcommand{\non}{\nonumber}

\newcommand{\rf}[1]{(\ref{#1})}
\newcommand{\ra}{\rightarrow}

\title{k-String Tensions and Center Vortices at Large N}

\author{Jeff Greensite \\ 
Physics and Astronomy Dept., San Francisco State University,
San Francisco, CA 94117 USA.
E-mail: \email{greensit@quark.sfsu.edu}}

\author{{\v S}tefan Olejn\'{\i}k \\ Institute of Physics, Slovak Academy 
of Sciences, SK-842 28 Bratislava, Slovakia. 
E-mail: \email{fyziolej@savba.sk}}

\abstract{We point out that there is a natural explanation, in terms of 
the center vortex confinement mechanism, for
the expected Casimir/Sine Law scaling of $k$-string tensions in the
large $N$ limit.  The crucial ingredient  is the existence of 
$Z_N$ center monopoles, which go over to U(1) monopoles in this limit.
Vortex densities leading to Casimir/Sine Law scaling at large-N
are constructed; these densities have no obvious pathologies and in 
particular do not grow with $N$.  We also note that center vortices are stable 
classical solutions of the Wilson action, for
all SU(N) gauge theories with $N>4$, and extend this old result to a 
broad class of lattice actions motivated by the improved action program
and the renormalization group.}


\preprint

\begin{document}

\section{Introduction}

   The term ``Casimir scaling'' was introduced by the present authors,
together with Del Debbio and Faber in ref.\ \cite{Cas}, as a
criterion for the validity of various confinement mechanisms.  Casimir
scaling refers to the fact, demonstrated 
numerically in refs.\ \cite{Cas1,Cas2},
that there is an intermediate distance interval (up to the onset of
color screening) where the string tension due to static color sources
in color group representation $r$ is approximately proportional to the
quadratic Casimir $C_r$ of the representation, i.e.
\beq 
         \s_r \propto C_r 
\eeq 
This representation dependence is seen quite convincingly
in data for the SU(3) group obtained by Bali and by Deldar
in refs.\ \cite{Cas2}.
Beyond the Casimir scaling regime, higher color charges are
screened by gluons, and the SU(N) asymptotic
string tension depends only on the $Z_N$ transformation properties
(or ``N-ality'') of the representation.  A correct theory of
confinement should be able to explain both Casimir scaling at
intermediate distances, where in particular there exists a non-zero
adjoint string tension, and the N-ality dependence of the string
tension at large distances, where the adjoint string tension vanishes.

   For SU(N) gauge theories with $N>3$ there are a number of color
representations, apart from the defining representation and its
conjugate, in which color charge cannot be screened by gluons down to
charge in a lower dimensional representation.  These unscreenable
representations each correspond to the lowest dimensional SU(N)
representation with N-ality $k$, represented by a Young tableau with
one column of $k$ boxes.  Because color screening cannot occur, it is
possible that Casimir scaling for these ``$k$-representations'' holds
asymptotically, and not just in an intermediate range.  If so, and if
Casimir scaling were exact, the Casimir scaling prediction is that
\beq
      {\s_k \over \s_1} = {k (N-k) \over N-1}
\eeq
An alternative scaling law proposed for $k$-string tensions is the
``Sine Law''
\beq
        {\s_k \over \s_1} = {\sin{\pi k/N} \over \sin{\pi/N}}
\eeq
which is motivated by MQCD \cite{MQCD}.

   In fact these two scaling laws are quantitatively
not so very different, for $N=4,5,6$. 
Numerical studies indicate that the Sine Law 
fits the SU(N) string tension data better than the Casimir Law 
in D=4 dimensions, while in D=3 dimensions the 
Casimir Law is the better of the two \cite{Teper,Pisa}.
Neither formula is a perfect fit in both D=3 and D=4 dimensions.
Moreover, as recently pointed out by Auzzi and Konishi \cite{Konishi},
the Sine Law cannot be universal among all
confining SU(N) theories, since it is found that there are 
non-universal corrections
to this law, for softly broken ${\cal N}=2$ supersymmetric gauge
theories.  

    In the large $N$ limit, however, Casimir scaling is exact in SU(N)
gauge theories due to the well-known factorization property.  In this limit
there is no difference at all between the Sine and Casimir scaling
laws for the $k$-representations, for $k$ finite as $N \ra \infty$.  
In general, for $k$-string tensions with $k \ll N$, 
we have from either law the simple prediction that
\beq
       {\s_k \over \s_1} = k  ~~~~~~~~~(k \ll N)
\label{ks}
\eeq
We will refer to this limiting behavior of Casimir scaling as ``$k$-scaling''.
It should be stressed that since $k$-scaling follows
from large $N$ factorization, this scaling law is guaranteed
to hold for large $N$ at intermediate distances (i.e.\ up to color screening),
and perhaps also asymptotically.

    The $k$-scaling law, because it may apply at asymptotic distance scales,
poses an interesting challenge to the center vortex theory of
confinement.  Let us consider SU(N) gauge theory at a very large but finite
value of $N$.  The vortex theory leads naturally to the conclusion that
the asymptotic string tension depends only on N-ality (i.e.\ on $k$,
for $k$-string tensions), and we have suggested in previous work with
Faber \cite{Us1} that Casimir scaling at intermediate distances is due
to the finite thickness of vortices.
$k$-scaling, however, may hold at arbitrarily
large quark separations, and if so it cannot be attributed to the structure of
the vortex core.  The interesting question is then whether $k$-scaling
is somehow implicit in the center vortex confinement mechanism.

   In this article we would like to offer the following
observation: The distribution of center vortices in an SU(N) gauge
theory is very likely controlled by an effective $Z_N$ gauge theory,
for reasons explained below.  This theory is quasi-local at the
color-screening scale.  As $N\ra \infty$, the $Z_N$ theory goes over
to a compact U(1) gauge theory, and a $Z_N$ Wilson loop of N-ality $k$
corresponds to the Wilson loop of an object with $k$ units of abelian
charge.  The charge-dependence of string tensions in a U(1) gauge
theory, in D=3 dimensions, was worked out some time ago by Ambj{\o}rn
and Greensite \cite{j2};  the same charge dependence is
found in $Z_N$ and U(1) lattice gauge theory at strong couplings in any
dimension greater than $D=2$.  
The dependence is precisely $k$-scaling, and is due to the formation of
$k$ distinct flux tubes forming between the $k$ quark and antiquark.  
Because of center dominance,
the string tension of $k$-charged loops in the effective U(1) theory is
identical to the string tension of $k$-strings in the corresponding
SU(N) theory at large $N$.  In this rather simple way, $k$-scaling at
large $N$ emerges quite naturally from the vortex mechanism, at least in
D=3, and probably also in D=4 dimensions.

   The following sections will elaborate on this observation, noting
in particular the relevance of center monopoles to the dynamics at
large $N$.  We will recall and extend a little-known fact,
pointed out many years ago by Bachas and Dashen \cite{BD}, that thin
center vortices are stable classical solutions of the Wilson SU(N)
lattice action for any $N>4$.  This fact, when extended to
effective and renormalization-group improved actions, is potentially
of great interest in the context of the vortex confinement mechanism.
Finally, we consider vortex densities in the large $N$ limit.
It is shown that vortex densities leading to $k$-scaling are well-behaved
in this limit and, contrary to some recent work \cite{Mitya2}
based on a dilute gas
approximation, do not grow with $N$.

\section{The Center Monopole Gas at Large N}

  Our confidence that the effective vortex theory is a $Z_N$ gauge
theory of some kind is based on extensive numerical computations of
center-projected observables in SU(2) gauge theory \cite{Us,Tubby,deF,dlcg}.  
Generalizing from our experience
with SU(2), we suggest that there is an effective $Z_N$ theory 
which is obtained
from the original SU(N) theory in the following way: Begin by fixing
the SU(N) lattice theory to an adjoint gauge, where the gauge-fixing
condition depends only on the link variables in the adjoint
representation.  Such adjoint gauges preserve a remnant $Z_N$
symmetry; examples include all the (direct, indirect, \dots) variants
of maximal center and laplacian center gauges.  Each link variable
$U_\m$ can be expressed as the product $U_\m = z_\m V_\m$,
where $z_\m \in Z_N$, and $z_\m I_N$ is the 
closest center element to $U_\m$ on the group manifold.
We can then write
\beq
      \int dU = \sum_{z\in Z_N} \int_{\cal R} dV
\eeq
where the domain ${\cal R}$ on the SU(N) manifold consists of
group elements $V$ satisfying
\beq
       \mbox{ReTr}[V] > \mbox{ReTr}[z V]  ~~~\mbox{for all~} z \in Z_N \ne 1
\eeq
The effective $Z_N$ theory is then formally
defined by integrating over $V$, i.e.
\beq
      \exp[-S_{eff}(z_\m)] \equiv \int_{\cal R} DV \d\Bigl[F[U_A]\Bigr]
            \Delta[U_A] e^{-S[zV]} 
\label{Seff}
\eeq
where the gauge-fixing condition applies to the link configuration $U_A$ 
in the adjoint representation.  It is not necessary to know $S_{eff}$
explicitly in order to simulate the effective theory numerically.
Evaluation of a $Z_N$ Wilson loop in the theory
defined by $S_{eff}(z)$ is equivalent to
the evaluation of a center-projected Wilson
loop in SU(N) lattice gauge theory, fixed to an adjoint gauge.

   There are some general arguments \cite{Yamada}, but no guarantee,
that center projection in adjoint gauges will locate confining center
vortices.  For this property to hold, the center projection must pass
(at least) three tests, concerning
\begin{enumerate}
\item  {\bf Vortex-limited Wilson loops $-$ } Let $W_{(n)}(C)$ denote
a Wilson loop in the defining representation of SU(N) gauge theory,
evaluated in the subset of configurations satisfying the constraint that
in each corresponding center projected configuration, the $Z_N$ loop
takes on the value
\beq
        Z(C) = \exp\Bigl[ 2i\pi n/N\Bigr]
\eeq
Then it is required that asymptotically, for large loops,
\beq
       {W_{(n)}(C)/W_{(0)}(C)} \ra \exp\Bigl[ 2i\pi n/N\Bigr]
\eeq
This condition is necessary if a projected loop actually identifies
vortices piercing the full loop.
\item {\bf Center Dominance $-$} ~ The string tension of projected loops
of a given N-ality in the adjoint gauge should equal the asymptotic string 
tension of the corresponding unprojected Wilson loops.  This condition
tells us that the center vortices located by the projection actually account
for the full confining force.
\item {\bf Vortex Removal $-$} ~ If all center vortices are removed
from each full lattice configuration, then the string tension should
vanish.  This should really be a consequence of the previous conditions,
and constitutes a check.  In practice, one checks whether the modified loops
\beq
      W'(C) = \langle \mbox{Tr}[V(C)] \rangle 
            = \langle Z(C) \mbox{Tr}[U(C)] \rangle
\eeq
have vanishing asymptotic string tension (where $U(C),~V(C),~Z(C)$ 
are products of the full link, V-link, and projected link variables,
respectively, around the contour $C$).
\end{enumerate}

   If all three conditions hold, then $S_{eff}(z)$ is the effective
action for the confining $Z_N$ flux.  The existence of such an
effective action is only known to be true from numerical experiments
in the case of SU(2) lattice gauge theory, in certain Laplacian center
gauges.\footnote{In maximal center gauges, center dominance holds good
only to an accuracy of about 20-30\% \cite{Borny}.  
In variants of Laplacian center
gauges, the accuracy is better than 10\% \cite{dlcg}.} 
Center dominance appears to
work for SU(3) in the original version of Laplacian center gauge
\cite{Pepe}; the
direct and indirect variants have not yet been investigated in the
SU(3) case.\footnote{There are also some SU(3) results in the direct
and indirect maximal center gauge, fixed by over-relaxation \cite{SU3}.}  
In the absence of evidence to the contrary, we will
assume that an effective $Z_N$ gauge theory satisfying the three
conditions above can be defined for \emph{any} SU(N) pure gauge theory
via an appropriate choice of adjoint gauge, as in the SU(2) case.

   The effective action $S_{eff}$ is certainly non-local at the lattice
scale, but it should be quasi-local at the color-screening scale.
This means that
the action can be expressed as a sum of loops (and loop products) on the
lattice, and the coefficients multiplying large loops in the action
must be exponentially suppressed with loop area, for loops whose extension
exceeds the color-screening scale.
If $S_{eff}$ were not quasi-local, it would be hard to understand how
large $Z_N$ Wilson loops could have an area law falloff, or how
the theory could avoid long-range correlations.

   The excitations of a $Z_N$ gauge theory are thin center vortices
and, for $N \ge 3$, center monopoles.  A center monopole is the
point (D=3) or line (D=4) of intersection of a number of vortices,
whose flux adds up to an integer multiple of $2\pi$ \cite{Yoneya}.  
For $N$ very
large, the $Z_N$ gauge group approximates a U(1) group, and center
monopoles go over to the usual abelian monopoles of compact QED.
Because the gauge group is compact, Dirac lines/sheets are invisible,
and isolated monopoles are stable saddlepoints of the effective action.
Since the action is non-local at the lattice scale, the action of
the monopole configurations is not set by the lattice cutoff, but
rather by the (color-screening) scale at which the effective theory
becomes quasi-local.
   
   Let us now consider an isolated monopole solution of the D=3 dimensional
U(1) effective 
theory.  The lattice Bianchi identities, coupled with spherical symmetry
at scales large compared to the lattice spacing, are sufficient to 
tell us that the field strength due to a monopole at the origin
approximates the continuum form far from the source
\beq
   F^{mon}_{\m\n}(x) = \e_{\m\n\a} {x^\a \over |x|^3} + F^{Dirac}_{\m\n} 
\label{mon}
\eeq
where $F^{Dirac}_{\m\n}$ is the contribution of the Dirac string.
A dilute gas of $n_m$ monopoles ($q=1$) and antimonopoles ($q=-1$)
is a superposition of such terms
\beq
       F_{\m\n}(x) = \sum_{a=1}^{n_m} q_a \e_{\m\n\a} {(x-x_a)^\a \over 
              |x-x_a|^3} + F^{Dirac}_{\m\n}(x)
\label{super}
\eeq
The field strength is expressed in terms of variables over a 
compact range
\bea
      F_{\m\n}(x) &=& A_\n(x+\widehat{\m}) - A_\n(x)
                    - A_\m(x+\widehat{\n}) + A_\m(x)
                  ~~~~~ -\pi \le A_{\a} \le \pi
\non \\
                  &=& \widetilde{F}_{\m\n} + 2\pi n ~~~~\mbox{where} ~~~~
          -\pi \le \widetilde{F}_{\m\n} \le \pi
\eea
Since $S_{eff}$ is compact and U(1) invariant, it depends on
$\widetilde{F}_{\m\n}$ rather than $F_{\m\n}$,
and Dirac lines contribute nothing to the action.
Moreover, if the effective U(1) theory is local at some scale, then
a weak-field low-wavelength expansion of the U(1) action
in powers of the field strength and its derivatives starts out
with the quadratic term, i.e.
\beq
 S_{eff} = \mbox{const.} + {1\over e^2} \sum_x \widetilde{F}_{\m\n}^2(x)
\eeq
Following Polyakov \cite{Polyakov}, and substituting the superposition
\rf{super} into the weak-field approximation for the effective action,
leads to Coulombic interactions between monopoles; i.e.\ to a monopole
Coulomb gas.  It should be noted that for a $Z_N$ theory with $N$
finite, the center monopole field cannot have the form \rf{mon}
arbitrarily far from the source, since there is a lower limit to the
non-zero $Z_N$ flux through a plaquette.  In a plasma, however, the
field of a monopole is screened at the Debye length, and the monopole
Coulomb gas analysis should be valid providing the U(1)
approximation \rf{mon} to the $Z_N$ monopole field holds at least up
to the screening length.

  It was shown in ref.\ \cite{j2} that a large Wilson loop with $k$
units of abelian charge, in a monopole Coulomb gas, has an area-law
falloff with string tension equal to $k$ times the string tension of a
single-charged loop.  By the center dominance property, the string
tension of loops in the effective U(1) theory are the same as the asymptotic
string tension of loops of the same U(1) charge (i.e.\ N-ality for
finite $N$) in the full SU(N) theory.  In this way we obtain the $k$-scaling
property in SU(N) in $D=3$ dimensions.

\section{Classical Stability of Center Vortices at $N>4$}

   Before turning to the density distribution of vortices in the
large-N limit, we would like first to discuss their classical stability
in the context of lattice gauge theory.
   
   In the continuum, instantons are stable local minima of the 
action of pure SU(N) gauge theory.  It is a remarkable but 
little-known fact, 
pointed out by Bachas and Dashen \cite{BD} in 1982, that an
analogous result holds in lattice theory: Thin
center vortices are stable local minima of the Wilson lattice
action for any SU(N) gauge group with $N>4$.  The proof is quite
trivial:  Any ``thin'' center vortex configuration on a classical vacuum
background is gauge equivalent,
in an SU(N) lattice gauge theory, to a lattice configuration in which
each link is a center element, i.e.
\bea
        U_\m(x) &=& Z_\m(x) I_N
\non \\
        Z_\m(x) &=& \exp\left[{2\pi i n_\m(x)\over N}\right]
         ~~~~~~~ (n_\m(x) = 1,2,...,N-1)
\label{thin}
\eea
It is not hard to see that any vortex-creating singular gauge 
transformation, operating on the classical vacuum, will result in
a configuration which can be transformed into this form.  
If any plaquette variable is different from unity, then it has been
pierced by a vortex.  Now consider
a small deformation of such thin vortex configurations
\bea
        U_\m(x) &=& Z_\m(x) V_\m(x)
\non \\
        V_\m(x) &=& e^{i A_\m(x)} ~~~,~~~~~ \mbox{Tr}[A^2_\m(x)] \ll 1
\label{deform}
\eea
The thin vortex is a classical solution if the Wilson action is
a (local) minimum for $V_\m(x)$ gauge equivalent to $V_\m(x)=I_N$.
Substituting \rf{deform} into the Wilson action for SU(N) gauge
theory
\beq
      S = {\b \over 2N} \sum_{p} \Bigl( 2N - \mbox{Tr}[U_p] - 
                                 \mbox{Tr}[U_p^\dg] \Bigr)
\eeq 
with $\b=2N/g^2$, we obtain
\beq
      S = {\b \over 2N} \sum_{p} \Bigl( 2N - Z_p \mbox{Tr}[V_p] - 
                         Z^*_p \mbox{Tr}[V_p^\dg] \Bigr)  
\eeq
Writing the product of $V$-link variables around a plaquette
in the usual way, as the exponential of a field strength
\beq
      V_p = \exp[iF_p] = I_N + i F_p - \oh F^2_p + ...
\eeq
and
\beq
       Z_p = e^{2\pi i n_p/N}
\eeq
we get, to the leading order in the field strength $F_p$ of the
deformation,
\beq
 S = {\b \over 2N} \sum_{p} \left[ 
     2N\left( 1- \cos\left({2\pi n_p\over N}\right) \right)  
     + \cos\left({2\pi n_p\over N}\right) \mbox{Tr}[F_p^2] + O(F^3) \right]
\eeq
From this expression it is clear that the action is minimized at
\beq
         \mbox{Tr}[F_p^2] = 0
\eeq
and hence that the vortex configuration \rf{thin} is stable,
\emph{providing} that at each plaquette with $n_p>0$, the condition
\beq
          \cos\left({2\pi n_p\over N}\right) > 0
\label{condition}
\eeq
is satisfied.
Otherwise, the vortex corresponds to a local maximum.
The above condition for vortex stability is satisfied iff
\beq
       {n_p \over N} < {1\over 4} \mbox{~~or~~} 
      {N-n_p \over N} < {1\over 4}
\eeq
For $N=2$ and $N=3$, which are the most studied cases, the stability
condition cannot be satisfied, and center vortices are clearly unstable
at the classical level.  Beginning, however,  with $n_p=1,4$ at $N=5$, 
vortex stability is obtained already from the classical action.

    This simple result can of course be extended beyond the simple Wilson
action, and therein lies its physical relevance.  It is obvious
that thin vortices, stable or not, are of no real importance at weak
couplings, because they are suppressed by a factor of order
\beq
       \exp\left[- {\mbox{Vortex Area} \over g^2} \right]
\eeq
and do not percolate.  The configurations which are of physical interest
are center vortices having some finite thickness in physical units.
In order to investigate the stability of vortices of thickness $d$ (or
of lesser thickness, but including quantum fluctuations up to scale $d$)  
starting from a lattice action at spacing $a$, we imagine following
the renormalization group approach, successively applying blocking
transformations of the form
\beq
       e^{-S'[\U]} = \int DU ~ \d[\U -F(U)] e^{-S[U]}
\eeq
where $\U$ are links on the blocked lattice, and 
$F(U)$ is a blocking function. The transformations are repeated until
a lattice action with lattice spacing $d$ is obtained.

    The Monte Carlo Renormalization Group (MCRG), and the closely
related perfect action approach, aim to compute
effective lattice actions at large scales.  It is assumed that
only a few contours (plaquettes, 6-link loops, 8-link loops...)
are important.  Given an effective action at a given length scale $d$,
we can then ask whether there exist stable center vortex solutions.
Let us consider, for simplicity, the class of improved lattice actions which
consist of plaquette plus $1\times 2$ rectangle terms, i.e.
\beq
   S = c_0 \sum_{plaq~P} (N-\mbox{ReTr}[U(P)]) + c_1 \sum_{rec~R}
                              (N-\mbox{ReTr}[U(R)])
\label{twopar}
\eeq
This class of actions has been widely discussed in the lattice 
literature, and includes the tadpole-improved action \cite{tadi},
the Iwasaki action \cite{Iwasaki}, and two-parameter approximations
\cite{QCD-TARO} to the Symanzik action \cite{Symanzik},
and the DBW2 action \cite{DBW2}.  For 
$c_0,~c_1 > 0$,
the stability of center vortices is trivial at $N>4$, 
and the analysis is exactly like the Wilson action case.
However, most improved actions of this type have $c_1 < 0$,
and so stability must be reconsidered. 

     We will now show that the condition for the stability
of of the trivial vacuum $U_\m(x)=I_N$ in the two parameter action
\rf{twopar}, which is
\beq
      c_0 + 8 c_1 > 0
\label{stab}
\eeq
is also sufficient to guarantee that thin vortices
satisfying eq.\ \rf{condition} are local mimina
of the two-parameter action, and therefore classically stable.
This stability condition is satisfied by all of the two-parameter
improved actions \cite{tadi,Iwasaki,QCD-TARO,Symanzik,DBW2}
mentioned above.

    It is enough to consider whether the action of a trivial vacuum
or vortex configuration can be lowered in a plane by some deformation
$V_\m(x) \ne I_N$ in eq.\ \rf{deform}. If this cannot be done in
a plane, where the Bianchi identity can be ignored and plaquette
variables chosen independently to minimize the action, then the 
action cannot be lowered in a volume either, since the action in
a volume is just the action in a set of planes.  Then, since we
are considering the case that $c_1$ is negative, we want to
restrict our attention to configurations in which the rectangle
contributions in eq. \rf{twopar} are as large as they can possibly
be, for a specified set of plaquette terms $N-\mbox{ReTr}[U_p]$
in the plane.  Consider any rectangle R containing plaquettes
$p_1,~p_2$, and write
\bea
      U(p_1) &=& Z(p_1) \exp[if_1 \widehat{e}_1 \cdot \vec{L}]
\non \\
      &=& Z(p_1) \Bigl(I_N + i f_1 e_1^a L_a
       - \oh f_1^2 e_1^a e_1^b L_a L_b + \dots \Bigr)
\non \\
      U(p_2) &=& Z(p_2) \exp[if_2 \widehat{e}_2 \cdot \vec{L}]
\non\\
      &=& Z(p_2) \Bigl(I_N + i f_2 e_2^a L_a
       - \oh f_2^2 e_2^a e_2^b L_a L_b + \dots \Bigr)
\eea
where $\widehat{e}_{1,2}$ are unit vectors, and $f_1,f_2$ are
positive.  Then the rectangle term has the form
\beq
      U(R) = U(p_1)gU(p_2)g^\dg
\eeq
where $g$ is an SU(N) group element corresponding to a certain
link variable on the rectangle, and in general
\beq
      gU(p_2)g^\dg = Z(p_2) \Bigl(I_N + i f_2 e_3^a L_a
       - \oh f_2^2 e_3^a e_3^b L_a L_b + \dots \Bigr)
\eeq
The rectangle variable, to second order in $f_1,~f_2$, is
\bea
  N &-& \mbox{ReTr}[U(R)] = 
    N(1- \mbox{Re}[Z(p_1)Z(p_2)]) 
\non \\ 
  &+& {1\over 4}\mbox{Re}[Z(p_1)Z(p_2)]
     \Bigl( f_1^2 + f_2^2 + 2 f_1 f_2 \widehat{e}_1 \cdot \widehat{e}_3 
            + \dots \Bigr)
\eea
and this is clearly as large as possible, for given $f_1,f_2 > 0$
and $\mbox{Re}[Z(p_1)Z(p_2)] > 0$, when
$\widehat{e}_1=\widehat{e}_3$.  The latter condition will always be satisfied
if we choose all link variables in the plane to lie in the same U(1)
subgroup of SU(N), so it is sufficient, for the purpose of
proving stability, to make this choice.  

    We consider small deformations around a one-vortex configuration,
which pierces plane at plaquette $p_0$, and denote the plaquettes
adjacent to $p_0$, as $\{p_i,~i=1-4\}$.  Writing again
that the deformation around a plaquette is
\beq
        V(p) = \exp[i f(p) \widehat{e} \cdot L]
\eeq
and defining
\bea
        r &\equiv& -{c_1 \over c_0}
\non \\
        z &\equiv& \cos\left({2\pi n_{p_0} \over N}\right)
\eea
we find the action in the plane $S_{plane}$, up to second order
in the deformation, to be
\bea
       Q &\equiv& {4\over c_0}\Bigl(S_{plane} - N(1-z)(1-4r)\Bigr)
\non \\
         &=& \sum_p f^2(p) - r \sum_p \sum_{\m=1}^2 
     (f(p) + f(p+\widehat{\m}))^2 
\non \\
         & & + (1-z) r \sum_{i=1}^4 
     (f(p_0) + f(p_i))^2  ~ - ~ (1-z)f^2(p_0)
\eea
where $p+\widehat{\m}$ denotes the plaquette adjacent to $p$ in
the $\widehat{\m} = \widehat{1} ~\mbox{or}~ \widehat{2}$  directions
tangent to the plane.  A little
rearrangement brings this to the form
\bea
   Q &=& \sum_p (1-8r) f^2(p) + r \sum_p \sum_\m (f(p+\widehat{\m})
        - f(p))^2 
\non \\
     & & +  (1-z) r \sum_{i=1}^4 
     (f(p_0) + f(p_i))^2  ~ - ~ (1-z)f^2(p_0)
\eea

   Now if $z=1$, so the expansion is around the trivial vacuum, then
we can immediately read off that $Q$ is minimized at $f(p)=0$,
and the trivial vacuum is stable, providing that $8r<1$, which is
the condition \rf{stab} above.  If, on the other hand, $8r>1$,
then $Q$ has no minimum to second order in the deformation, and
the trivial vacuum is unstable.  

   Next, separating out of the first two sums in $Q$ the contributions
containing the plaquette $p_0$, we have
\bea
     Q &=& \sum_{p \ne p_0} (1-8r)f^2(p) + r \sum_{p,p+\widehat{\m} \ne p_0}
         (f(p+\widehat{\m}) - f(p))^2 
\non \\ 
    & &  + r \sum_{i=1}^4  (f(p_i) - f(p_0))^2
         + r(1-z) \sum_{i=1}^4  (f(p_i) + f(p_0))^2
         + (z-8r) f^2(p_0)
\non \\
      &=& \sum_{p \ne p_0} (1-8r)f^2(p) + r \sum_{p,p+\widehat{\m} \ne p_0}
         (f(p+\widehat{\m}) - f(p))^2
\non \\ 
    & &  + rz \sum_{i=1}^4  (f(p_i) - f(p_0))^2
         + 2r(1-z) \sum_{i=1}^4  f^2(p_i)
         + z(1-8r) f^2(p_0)
\eea
If $z>0$, which was the stability condition for vortices in the Wilson
action, and if $r<1/8$, which is the stability condition for the
trivial vacuum in the two-parameter action, then every term
contributing to $Q$ is positive semi-definite.  The minimum $Q=0$ is
only obtained at $f(p)=0$, i.e.\ at $V(p)=I_N$ everywhere, and the
vortex is therefore a stable local minimum of the two-parameter
action.

   So it appears that the result first obtained by Bachas and Dashen
is quite robust: center vortices at $N>4$ are 
stable minima 
of lattice actions in a large region of coupling-constant space
associated with improved actions, and it seems unlikely that
adding a few more contours, as in certain proposed 
perfect actions \cite{deGrand,Niedermeyer}, would alter this
result.\footnote{We have not investigated those actions
systematically, however.}  If in fact center vortices remain as local minima
of the action all along the renormalization trajectory, then
their physical effects must become apparent at some scale.
The argument is simply that
the Boltzmann suppression factor 
of a vortex in an effective (or perfect) action at
lattice spacing $d$ goes like
\beq
        \exp\left[ -{\mbox{Vortex Area}\over \k^2(d)} \right]
\eeq
where the vortex area is in lattice units. 
On the other hand, the entropy factor increases with 
vortex surface area as
\beq
        \exp\left[ + c \times \mbox{Vortex Area}  \right]   
\eeq
where $c$ is a constant.  As $d$ increases, $\k^2(d)$
also increases. Eventually entropy wins over action, and vortices
at that scale will percolate through the lattice.

   So far it would appear that vortices are \emph{only} stable
at $N>4$.  At this point we will digress briefly to consider what happens for
$N=2,3,4$, following closely the discussion in ref.\ \cite{scup}.  
The salient point is that 
the effective long-range action at the color-screening scale 
should contain contours in the adjoint representation, multiplied by 
coefficients falling with a perimeter rather than an area law. 
The reasons for this are easiest to see in the context of the
strong-coupling expansion.
Let us start with the strong-coupling Wilson action with lattice
spacing $a$, and construct a blocked action with lattice
spacing $La$
\beq
       e^{-S'[\U]} = \int DU ~ \prod_{l'} 
           \d[\U_{l'}-(UUU...U)_{l'}] e^{-S_W[U]}
\eeq
One readily finds the leading terms in the blocked action
\bea
     -S'[\U] &=& 
  \sum_{p'} \left\{ 2 N\left({1\over g^2 N}\right)^{L^2}
  \mbox{ReTr}_F [\U(p')] \right. 
\non \\ 
  & & \left.
 + 4(D-2) \left({1\over g^2 N}\right)^{4(4L-4)}
  \mbox{Tr}_A [\U(p'] \right\} 
\non \\
  & & + \mbox{larger adjoint loops with $e^{-P}$ coefficients}
\eea
The adjoint loops are non-planar contributions, and arise from the 
``tube'' diagrams shown in Fig.\ \ref{tube}.  The larger adjoint loops
are important for the color-screening of zero N-ality Wilson loops,
and cannot be ignored if accurate results for such loops are required. 

\FIGURE[t]{
\centerline{\scalebox{0.60}{\includegraphics{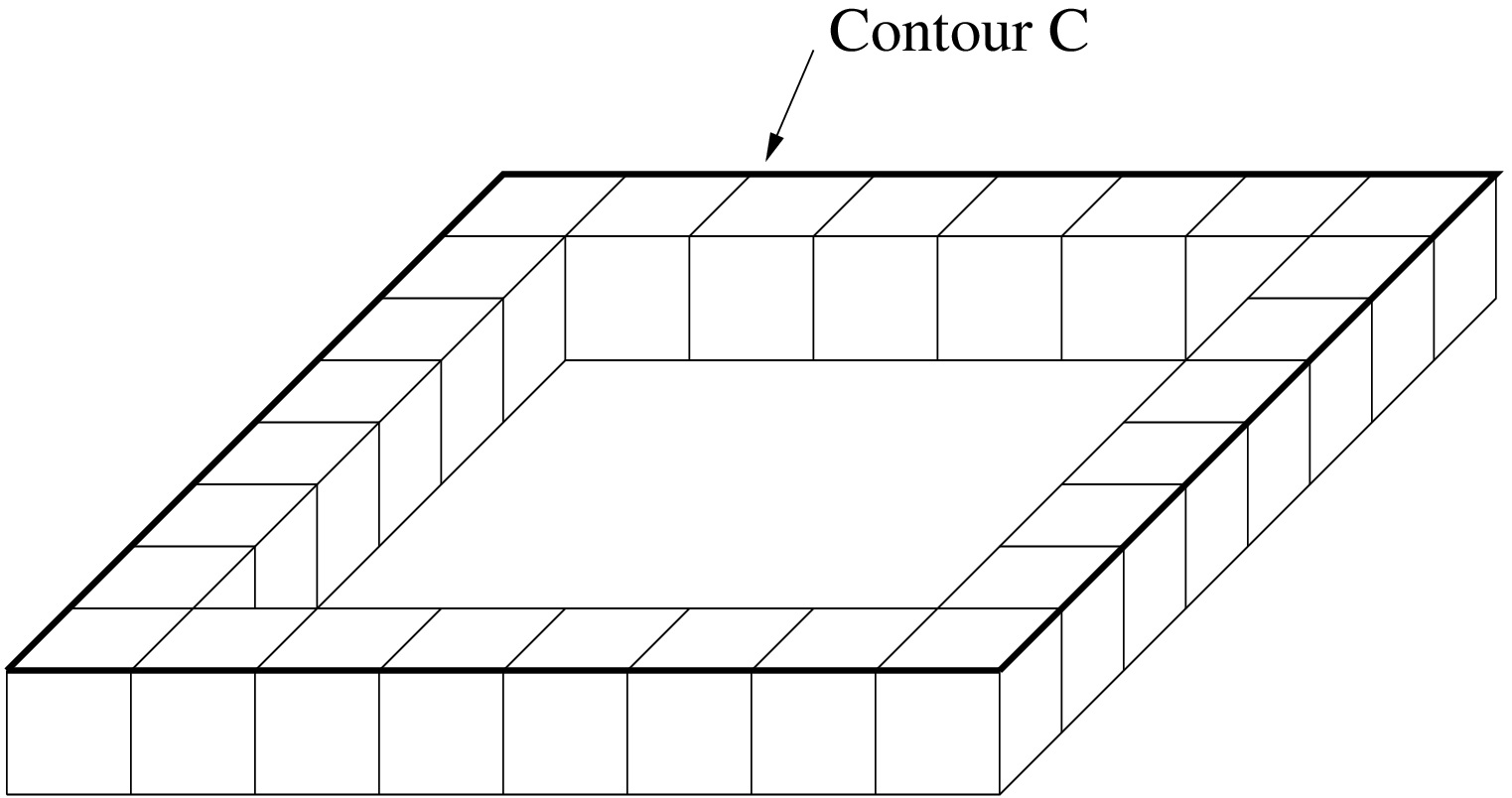}}}
\caption{Diagrams leading to perimeter-law falloff, in the
blocked action, of adjoint-representation loops.}
\label{tube}
}

   In fact the non-local blocked action can be expressed 
(with some additional complications) 
as a local action containing a number of adjoint Higgs fields \cite{scup}.
For simplicity, however, we will simply consider the above action 
truncated to the leading plaquette terms, i.e. 
\beq
   -S'_{trunc}[\U] = \sum_{p'} \Bigl\{ c_0 \mbox{ReTr}_F[\U(p')] +
                                   c_1 \mbox{Tr}_A[\U(p')] \Bigr\}
\eeq
\ni with 
\beq
       c_0 \sim \exp[-\s \mbox{~Area}(p')]  ~~~,~~~
   c_1 \sim \exp[-4 \s \mbox{~Perimeter}(p')]   
\eeq
Consider small fluctuations around a vortex configuration
\beq
 S'_{trunc} = \mbox{const} + {1\over 2} \sum_{p'} \left\{ 
       c_0 \cos\left({2\pi n_{p'}\over N}\right) \mbox{Tr}_F[F_{p'}^2] 
     + c_1 \mbox{Tr}_A[F_{p'}^2] \right\}
\eeq
It is clear that at large blocking $L$, $c_1$ is much greater than
$c_0$. Then $S'_{trunc}$, at the scale where $c_1 \gg c_0$, is
minimized at $F_{p'}=0$.  This means that at this scale, vortex
configurations are stable even for $N=2,3,4$.\footnote{We would like
to comment at this point on an argument against vortex stability
in Yang-Mills theory
(i.e.\ against any definite thickness for vortices) put forward in
ref.\ \cite{Terry}, and based on an inequality involving a certain
vortex operator.  We wish to point out that the argument of that
reference could be just as well applied to $Z_2$ lattice gauge theory
at strong couplings, arriving at the same inequality as in the
Yang-Mills case, and reaching the false conclusion that $Z_2$ vortices
also have no definite thickness.  We believe that the problematic
aspect of that analysis was to identify a system in which vortices are
restricted to a fixed set of vortex containers, associated with the
derived inequality, with the actual vortex vacuum.  The restriction of
vortices to lie in containers at fixed locations would be, for the
vortex vacuum, a drastic truncation of the entropy of vortices due to
positional fluctuations.  In fact, numerical experiments with both
$Z_2$ lattice gauge theory, and center-projected lattices in
Yang-Mills theory, have shown that confinement is due to a single
vortex configuration which percolates through the entire lattice.  The
vortex container of such a configuration has a volume comparable to
the volume of the full lattice, but this fact in no way implies that
the thickness of the vortex is the length of the lattice.}

   This result was derived for the strong coupling Wilson action.
However, from the renormalization-group point of view, the perfect
action at the color screening scale (which is approximately one fermi
for $N=2$) is surely a strong-coupling action of some sort, and the
same analysis should apply.  There is also an
interesting study of vortex stability in the framework of continuum
QCD, that has been carried out recently
by Diakonov and Maul \cite{Mitya1}.

\section{Density of Vortices at Large N}

    Numerical simulations in SU(2) lattice gauge theory indicate
that a dilute gas approximation for vortices is not really valid 
in the case of $N=2$ colors \cite{PGL}, and a ``vortex liquid''
picture is probably more appropriate.  
The limitations of the dilute gas approximation,
as we will see, are even more apparent at large $N$.

    Following the notation of Del Debbio and Diakonov \cite{Mitya2},
define an ``$l$-vortex'' as a center vortex which, when it pierces
the minimal area of a Wilson loop, contributes a factor 
$z_N^l=\exp[2\pi il/N]$.  A vortex which pierces the loop area twice
in opposite directions, and is therefore
not topologically linked to the loop, contributes the trivial factor
$z_N^l z_N^{N-l} = 1$.
The dilute gas approximation assumes that: (i) vortex piercings are
statistically independent; (ii) there is no limitation on the number
of vortices that can pierce a given area; (iii) it is always possible to
distinguish between piercings in a plane due to a single 
$l$-vortex, and piercings due to, e.g, $l$ 1-vortices.  With these 
assumptions, the
probability that $n_l$ $l$-vortices pierce a Wilson loop is given
by the Poisson distribution
\beq
         P_{n_l} = {\overline{n}_l^{n_l} \over n_l !}
                   e^{-\overline{n}_l}
\label{PD}
\eeq
where
\beq
        \overline{n}_l = \r(l,N) \mbox{Area}(C)
\eeq
is the average number of $l$-vortices piercing the minimal area
of the Wilson loop, and $\r(l,N)$ is the number of $l$-vortices
per unit area piercing any given plane.  In the dilute gas approximation,
the contribution of vortex piercings to an SU(N) Wilson loop 
in the $k$-representation is
\bea
      W(k,N) &=& \prod_{l=1}^{N-1} \sum_{n_l=0}^{\infty} P_{n_l}
                      (z_N)^{kln_l} 
\non \\
             &=& \exp[-\s^{PD}(k) \mbox{Area}(C)]
\label{dga}
\eea
with the $k$-string tension
\beq
      \s^{PD}(k) = \sum_{l=1}^{N-1} \r(l,N)\Bigl(1 - 
                   e^{2\pi i kl/N} \Bigr)
\label{dilute}
\label{spd}
\eeq
where $\s^{PD}(k)$ and Area$(C)$ are in physical units. The ``PD''
superscript is a reminder that this string tension is associated with
the Poisson distribution

   If one accepts the assumption of a dilute vortex gas, then it is 
interesting to ask what Casimir or Sine Law
scaling implies for the vortex density $\r(l,N)$.  This question
was addressed in a recent article by Del Debbio and Diakonov 
\cite{Mitya2}, who find that the scaling laws are obtained from
eq. \rf{dilute} by the vortex densities (for $l \ll N$)
\beq
      \r(l,N) = N\s(1) \times \left\{ \begin{array}{cl}
 {2\over \pi^2 (4l^2-1)} + O\left({1\over N^2}\right) & 
  \mbox{~~~Sine Law} \cr
          &   \cr
 {N\over 2(N-1)\pi^2 l^2} + O\left({1\over N^2}\right) & 
  \mbox{~~~Casimir Scaling} \end{array} \right.
\label{D3}
\eeq
whose leading terms are proportional to $N$.

   A vortex density growing linearly with $N$ is certainly
pathological, but in fact this behavior is not what is found
in center vortex theories having the $k$-scaling property.
We will first supply two examples, namely strong-coupling $Z_N$ lattice
gauge theory and compact $QED_3$, which will serve to illustrate this point.
Next, we will derive explicit center vortex densities leading to 
$k$-scaling, and show that these densities are proportional to
$1/N$, rather than $N$.  Finally, the apparent contradiction with the
result \rf{D3} is explained, by showing that finite-range correlations
in the center field strength, combined with the $k$-scaling property,
are inconsistent with the assumed Poisson distribution \rf{PD}, from which
eq.\ \rf{D3} is derived.

   We begin with strong-coupling $Z_N$ 
lattice gauge theory, with Wilson action
\bea
      S &=& -\b \sum_{p} \Bigr(Z(p) + Z^*(p)\Bigl)
\non \\
        &=&  -2\b \sum_{p} \cos\left[{2\pi n_p \over N}\right]
\eea
Let $A(C),P(C)$ denote the minimal area and perimeter of loop $C$
in lattice units.  For $D=2$ dimensions, to leading order in $\b$,
a Wilson loop in the N-ality $k$ representation (with
$k<N/2$) has a vacuum expectation value
\beq
      \langle Z^k(C) \rangle = \left({\b^k \over k!}\right)^{A(C)}
          ~~~~~ (D=2)
\label{2dzk}
\eeq
with corresponding string tension (in lattice units)
\beq
        \s_k = k\log(1/\b) + \log(k!) 
\label{2dst}
\eeq
which has $k$-scaling only for $k \ll \b$.  The result is different
for $D>2$ dimensions. To leading order in $\b$ we find instead
\beq
      \langle Z^k(C) \rangle = \b^{k A(C) + \m_k P(C)}
\label{4dzk}
\eeq
with a string tension
\beq
        \s_k = k\s_1 ~~,~~ \s_1 = \log(1/\b) ~~~~~~ (D>2) 
\label{4dst}
\eeq
which has perfect $k$-scaling for $k<N/2$.  The difference between
the $D=2$ and $D>2$ cases is that for $D=2$ it is necessary to
expand $\exp(\b Z(p))$ to $k$-th order in $\b$ for every plaquette
in the area bounded by loop $C$.  This brings in a factor of $(1/k!)$.
By contrast, in any number of dimensions 
greater than two, it is possible to bring down a single
layer of plaquettes on $k$ distinct horizontal surfaces, thereby expanding 
$\exp(\b Z(p))$ only to first order at each plaquette on each of
these surfaces.  A cross-section of the leading strong-coupling
diagram, for the case $k=5$, is shown in Fig.\ \ref{cross1}.  Overlapping
plaquettes at the boundaries account for the perimeter contribution
$\exp(-\m_k P(C))$ in eq.\ \rf{4dzk}.  

\FIGURE[t]{
\centerline{\scalebox{0.60}{\includegraphics{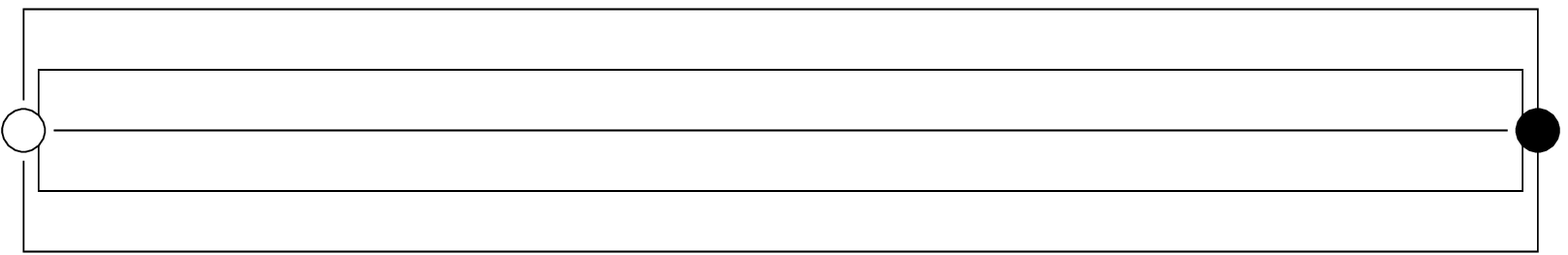}}}
\caption{A cross-section of the leading strong-coupling diagram,
in $D>2$ dimensions, for the expectation value of a $k=5$ Wilson
loop in strong-coupling $Z_N$ lattice gauge theory.  The loop
is perpendicular to the page, crossing it at the open and solid circles.
Each solid line is a cross-section of a surface of plaquettes, with 
surfaces partially overlapping at the right and left-hand boundaries.
In Hamiltonian formulation, the solid horizontal lines represent
$k$ distinct flux tubes running between a $k$-quark and antiquark.}
\label{cross1}
}

    What our simple example demonstrates is that $k$-scaling readily 
coexists with center vortex/center monopole confinement mechanisms.
Strong-coupling $Z_N$ lattice gauge theory is a theory with only
center vortex and (in $D>2$ dimensions) center monopole excitations;
there are no other physical degrees of freedom in the theory.
It is also a theory in which the string tension satisfies $k$-scaling
perfectly, for the $D>2$ case where center monopoles exist.  

   Let us consider the vortex density $\r(l,N)$ in strong-coupling
$Z_N$ lattice gauge theory, with lattice spacing $a$.  The quantity
$\tr(l,N)=\r(l,N)a^2$ is the probability that an $l$-vortex
pierces a plaquette.  As $\b \ra 0$ all plaquette values are
equally likely, i.e.\ $\tr \ra 1/N$ in this limit, and in general
\beq
   \r(l,N) = {1\over Na^2}\left[1 + O(\b) \right]
\label{rhoscup}
\eeq
at strong couplings.
Far from growing linearly with $N$, this density actually
\emph{falls} with $N$.  Here we have a complete contradiction
to the behavior \rf{D3} derived from the dilute gas approximation.
In the special case of 
$D=2$ dimensions, where plaquettes all fluctuate independently, the
density is obtained immediately from the Wilson action
\bea
      \r(l,N) &=& {1\over a^2} {\exp[2\b \cos(2\pi l/N)] \over
          \sum_{m=0}^{N-1} \exp[2\b \cos(2\pi m/N)] }
\non \\
         &\approx& {1\over Na^2 }
       { \exp[2\b \cos(2\pi l/N)] \over 1 + \b^2 }
\label{density}
\eea
which illustrates the typical $1/N$ dependence of vortex densities in
strong-coupling lattice gauge theory.

  The fact that strong-coupling $Z_N$
lattice gauge theory is a counter-example to the dilute gas result
for the vortex density, eq.\ \rf{D3}, is no doubt
due to the fact that the assumptions underlying the dilute gas
approximation are not satisfied by the $Z_N$ lattice gauge theory.
In particular, in $Z_N$ lattice theory, no more than one vortex
can pierce a plaquette.   
If a Wilson loop around a plaquette has a value
$\exp(2\pi i l/N)$ with $l\ne 0$, this means that one $l$-vortex 
has pierced the plaquette.  As a consequence, a planar area $A(C)$
in lattice units can be pierced by no more than $A(C)$ vortices.
There is no such restriction in the dilute gas approximation,
and the sum over $n_l$ in eq.\ \rf{dga} runs to $n_l=\infty$.

   A second example of $k$-scaling by center flux is provided by 
compact U(1) lattice gauge theory in $D=3$ dimensions.  As explained
in section 2, this example is
our paradigm for confinement by center degrees of
freedom at large $N$.  Once again, this is a theory in which 
center degrees of freedom are the only degrees of freedom in the 
theory, since the center of the gauge group is the gauge group itself.
In compact $QED_3$ there is no need to invoke the strong-coupling
expansion to demonstrate $k$-scaling; instead we can apply 
Polyakov's monopole Coulomb gas analysis to calculate $k$-string tensions.  
This analysis was carried out in ref.\ \cite{j2}, where it was found that  
$k$-scaling is obtained through the existence of 
$k$ independent surfaces of electric flux (i.e.\ $k$ separate flux tubes
at any given time).  As we have seen, this is the same mechanism which accounts
for $k$-scaling in strong-coupling $Z_N$ lattice gauge theory in 
$D>2$ dimensions. The formation of $k$ separate flux tubes also 
accounts for Casimir scaling, in the large $N$ limit, in the 
gluon-chain model \cite{gluonchain}. 

   The example of compact $QED_3$ also illustrates another important
point: The center confinement mechanism does not necessarily depend on
having center flux of some fixed magnitude concentrated in tubes (3D)
or sheets (4D).  Center flux in compact $QED_3$, passing through a
given area in a plane, can spread out through the volume in any manner
consistent with the U(1) lattice Bianchi identity; there is no reason
for a fixed quantity of center flux to remain concentrated in a tube
or a sheet-like region.  What \emph{is} essential to the center
confinement mechanism is that center flux passing through neighboring
regions of a plane, of sufficiently large area, are uncorrelated.  In
$Z_2$ lattice gauge theory, where the center flux cannot spread out,
the lack of correlation between center flux through neighboring
regions is accomplished by percolation of vortices through the entire
lattice.  In compact $QED_3$, where the center flux can spread out and
the flux through a plaquette can be arbitrarily small, disordering is
accomplished via a monopole plasma.  In $Z_N$ theories at moderate
$N$, presumably both vortex percolation \emph{and} center monopole
effects are in play.

   The short-range correlation of center flux is sufficient,
in any $Z_N$ or U(1) theory,  to derive the area law falloff of Wilson
loops.  Combined with center dominance, short-range center flux
correlation is also
is sufficient to derive the area law in SU(N) gauge theories.  
The derivation for 
$Z_N$ theories, including the effective P-vortex theory extracted
(via eq.\ \rf{Seff})
from SU(N) gauge theory, goes as follows:  We imagine dividing a plane
into square regions of area $\A$, with $\A$ taken large enough so 
that the center flux piercing different regions of the plane are uncorrelated.
Then consider a large rectangular loop 
$C$ in the plane, with minimal area Area$(C) \gg \A$, and subdivide this 
minimal area into Area$(C)/\A$ adjacent regions of area
$\A$ bounded by loops $\{C_i, i=1,...,\mbox{Area}(C)/\A\}$.  
In a $Z_N$ theory we have the identity
\beq
        Z^k(C) = \prod_i Z^k(C_i)
\eeq
Then, using the fact that the flux through the regions are uncorrelated,
we have
\bea
     \langle Z^k(C) \rangle &=& \langle \prod_i Z^k(C_i) \rangle
\non \\
       &=& \prod_i \langle Z^k(C_i) \rangle
\non \\
       &=& \langle Z^k(C_i) \rangle^{\mbox{\scriptsize Area}(C)/\A}
\eea
so that
\beq
       \s(k) = - {1\over \A} \ln \Bigl[\langle Z^k(C_i) \rangle \Bigr]
\eeq
where $\s(k)$ is the string tension in physical units.
Now we define $\tr(l,\A)$ to be the probability that a region
of area $\A$ is pierced by flux $2\pi l/N$.  We make no distinction
about how that flux is divided within the region (e.g.\ whether the
total flux is provided by two vortices of type 
$l/2$, or one vortex of type $l$, etc.). Then 
\beq
     \langle Z^k(C_i) \rangle = \sum_{l=0}^{N-1} \tr(l,\A) z_N^{kl}
\label{Zk}
\eeq
and the string tension is
\beq
   \s(k) = - {1\over \A} 
        \ln \left[ \sum_{l=0}^{N-1} \tr(l,\A) z_N^{kl} \right]
\label{sk}
\eeq

   If a $Z_N$ lattice gauge theory has both $k$-scaling and short-range
correlations in the flux through a plane $-$ and these conditions hold
for both strong-coupling $Z_N$ theory and compact $QED_3$ $-$ then we
can derive the corresponding $\tr(l,\A)$.  From eq.\ \rf{Zk}, we see
that $\tr(l,\A)$ is the inverse discrete Fourier transform of 
$\langle Z^k(C_i)\rangle$ 
\begin{equation}\label{IFT}
\tr(l,\A)=\frac{1}{N}\sum_{k=0}^{N-1}\langle Z^k(C_i)\rangle
    \;e^{-2\pi i k l/N}
\end{equation}
and inserting the $k$-scaling behavior
\begin{equation}
\langle Z^k(C)\rangle= \left\{ \begin{array}{cl}
          e^{-k\s(1) \A} & ~~k \le {N\over 2} \cr
          e^{-(N-k)\s(1) \A} & ~~k>{N\over 2} \end{array} \right.
\end{equation}
we find that
\bea
\tr(l,\A) &=&
  \frac{1}{N}\left\{
\frac{1-\gamma^{\left[\frac{N}{2}\right]+1}
e^{-2\pi i (\left[\frac{N}{2}\right]+1) l/N}}{1-\gamma e^{-2\pi i l/N}}+
\frac{1-\gamma^{\left[\frac{N}{2}\right]+1}
e^{2\pi i (\left[\frac{N}{2}\right]+1) l/N}}{1-\gamma e^{2\pi i l/N}} \right.
\non \\
 & &  \left. -1  - \d_{2\left[\frac{N}{2}\right],N} 
    (-1)^l \gamma^{\left[\frac{N}{2}\right]} \right\}
\non \\
 &=& \frac{1}{N} \left\{
\frac{(1-\gamma^2)
\left(1 - (-1)^l \gamma^{\left[\frac{N}{2}\right]}\right)}
{1 + \gamma^2 - 2 \gamma \cos\left(\frac{2 \pi l}{N}\right)}
\right\} ~~~~\mbox{for even-integer $N$}
\non \\
 &=& \frac{1}{N} \left\{
\frac{(1-\gamma)\left(1+ \gamma - 2 (-1)^l
\gamma^{\left[\frac{N}{2}\right]+1}
\cos\left(\frac{\pi l}{N}\right)\right)}
{1 + \gamma^2 - 2 \gamma \cos\left(\frac{2 \pi l}{N}\right)}
\right\} ~~~~\mbox{for odd-integer $N$}
\non \\
\label{trho}
\eea
where we have defined
\beq
       \gamma = e^{-\s(1) \A} ~~~\mbox{and}~~~
       \left[\frac{N}{2}\right] \equiv \left\{ \begin{array}{cl}
       {N\over 2} & \mbox{~~even~}N \cr
                  &                 \cr
     {N-1\over 2} & \mbox{~~odd~} N \end{array} \right.
\eeq
\ni It can be checked that for any $N$, and any $\s(1)>0$,
\beq
   \sum_{l=0}^{N-1} \tr(l,\A) = 1 ~~,~~ 
   \mbox{and} ~~~\tr(l,\A) > 0 ~~\mbox{for every~} l
\eeq
and one can also verify that in the $\s(1) \ra 0$ limit, 
only $\tr(0,\A) \ra 1$ is non-zero 
(i.e. no string tension means no center flux).
Once again we see, in eq.\ \rf{trho}, 
the overall factor of $1/N$ multiplying the flux
probability distribution, as opposed to an
overall factor of $N$ that would have been
expected from the dilute gas result $\rf{D3}$. 

   Equation \rf{trho} is a very general result for $Z_N$ gauge theories.  
It assumes only $k$-scaling and finite-range flux correlations.  The
result therefore holds for \emph{any} $Z_N$ theory with these properties, 
and these include
strong-coupling $Z_N$ gauge theory for $D>2$, compact $QED_3$ at any
coupling, and any other $Z_N$ theory (such as the complicated, non-local 
P-vortex theory defined in eq.\ \rf{Seff}) whose long-range structure is 
described by a monopole Coulomb gas.   Our result demonstrates quite
clearly that there is no incompatibility whatever between a center flux
confinement mechanism and $k$-scaling of the string tensions.

\subsection{Why Vortex Flux Cannot Follow a Poisson Distribution at Large $N$}

   The center flux probabilities of eq.\ \rf{trho}, and the vortex density 
derived from the Poisson distribution, eq.\ \rf{D3}, apparently
contradict one another.
The former expression is proportional to $1/N$ while the latter grows linearly
with $N$, and this discrepancy needs to be explained.  
We will now show that the combined properties of $k$-scaling, 
and the existence of finite-range correlations among center field strengths,
are inconsistent with the Poisson distribution \rf{PD} at large $N$, from which
eq.\ \rf{D3} is derived.

   The derivation of eq.\ \rf{trho} starts from the assumption that
one can subdivide the plane into squares of some sufficiently large
area $\A$, such that the center fluxes in each area are essentially
uncorrelated.  In every example we have suggested, where center flux
is responsible for confinement, there is a lower bound $\A_{min}=L_c^2$
to the area $\A$, where $L_c$ is the correlation length for the center
field strength.  In strong-coupling $Z_N$ lattice gauge theory,
$\A_{min} = a^2$, where $a$ is the lattice spacing, so in this case $L_c=a$.
In compact $QED_3$, $L_c$ is the average monopole
separation.  In center-projected QCD (the theory defined by $S_{eff}(z)$
in eq.\ \rf{Seff}),
$L_c$ must be on the order of the characteristic 
length scale of the theory, i.e.\
$L_c \sim O(\Lambda_{QCD}^{-1})$.  Below these length scales, all field
strength fluctuations are correlated, and a treatment of center flux
in terms of statistically independent vortex piercings is inconsistent.
Thus, to describe the center flux in a plane in terms of statistically
independent fluctuations, which is certainly a prerequisite
for the use of the Poisson distribution, it is necessary to follow
the procedure outlined above:  Sub-divide the plane into squares
of area $\A=\A_{min}$, and identify the center flux $2\pi l/N$ 
in a given square with a ``piercing'' of the square by a vortex 
of type $l$.\footnote{As explained above in connection with compact
$QED_3$, the term ``vortex'' in this context should not be taken to
mean that the center flux of type $l$ piercing a square region 
will necessarily remain 
concentrated  in a tube or sheet of constant flux $2\pi l/N$.}
The vortex density is then defined as
\beq
      \rho(l,N) = {\tr(l,\A) \over \A}
\label{vd}
\eeq
where $\tr(l,\A)$ is the probability that the center flux in the
square is $2\pi l/N$.
From eq.\ \rf{sk}, and the fact that the probabilities
$\tr(\l,\A)$ sum to unity, we have
\bea
       \s(k) &=& {-1\over \A} \log\left[\tr(0,\A) + 
         \sum_{l=1}^{N-1} \tr(l,\A) z_N^{kl} \right]
\non \\
            &=& {-1\over \A} \log\left[ 1 - 
     \sum_{l=1}^{N-1} \tr(l,\A) \left(1 - 
        \exp\left({2\pi ikl \over N}\right) \right)\right]
\non \\
           &=& {-1\over \A} \log\left[ 1 - \A \s^{PD}(k) \right]
\label{correct}
\eea
where $\s^{PD}(k)$ is the string tension 
in eq.\ \rf{spd} derived from the Poisson
distribution.  Eq.\ \rf{correct}
is the correct formula for the string tension $\s(k)$, for uncorrelated
center flux in areas $\A$.  However, the approach based on 
the Poisson distribution agrees with this correct
answer only if all the $\A \s^{PD}(k)$ satisfy
\beq
        \A \s^{PD}(k) \ll 1   ~~\mbox{all}~~k \in [0,N-1]
\eeq
We can now see that this condition is incompatible with the
$k$-scaling formula 
\beq
        \s(k) = k \s(1)  ~~~ \left( k < {N\over 2} \right) 
\eeq
at sufficiently large $N$, and in fact can hold only for 
\beq
        N  \ll {2 \over \A \s(1)}
\eeq

  We conclude that the following conditions
\begin{enumerate}
\item $k$-scaling;
\item finite correlation length $L_c \Rightarrow$ 
finite lower bound for $\A$;
\item Poisson distribution for vortex piercings;
\end{enumerate}
are incompatible at sufficiently large $N$.  This means that the
result of ref.\ \cite{Mitya2}, shown in eq.\ \rf{D3} above and indicating
that $k$-scaling requires a vortex density growing linearly with $N$,
is not valid for realistic theories with a finite range of center
flux correlations.   

   An even simpler argument is just to note that $\tr(0,\A) \sim 1/N$
for fixed $\A=\A_{min}$ at large $N$.  This means that each square region
is virtually certain to be pierced by some finite amount of center flux
(as, e.g., in the case of compact $QED_3$), thereby
violating the diluteness assumption underlying the Poisson distribution.

\section{Conclusions}

   The $k$-scaling of $k$-string tensions at large $N$ appears to be a natural
outcome of a confinement mechanism based on center degrees of freedom.
Although $k$-scaling, as the common large-N limit of Casimir and Sine Law
scaling, is certainly not unique to the center vortex scenario,
it is still of interest to see how the property emerges in that context.
Stated briefly, the confining dynamics at asymptotic distances
are described at large $N$ by a 
center monopole Coulomb gas.  The $k$-string tension, in that system,
is proportional to $k$, because there are $k$ independent flux tubes 
stretching between the quark and antiquark.
 
   Thin center vortex configurations 
are stable classical solutions of the SU(N) Wilson lattice
action for any $N>4$, as shown in ref.\ \cite{BD} and reviewed here.
We have extended this result to a class of two-parameter lattice actions
that are motivated by renormalization-group 
and tadpole-improvement considerations.
It is found that for all couplings such that the trivial vacuum is
a stable minimum of the two-parameter action, thin center vortices are also 
stable local minima of the action.  We conjecture that for $N>4$, 
thin center vortex configurations 
are stable minima of the perfect lattice action 
all along the renormalization trajectory.  If that is so, then vortex 
stability at the semiclassical level is not just a lattice artifact,
but is a genuine feature of the continuum theory.

   Finally we have shown, by explicit examples, that the $k$-scaling
property does not imply any pathological property of vortex densities
in the large-N limit, of the type that was suggested in 
ref.\ \cite{Mitya2}.  Those pathologies derive from the use
of a Poisson distribution for vortex densities, which we have shown here 
to be inconsistent with $k$-scaling and the existence of
a finite correlation length in the large-N limit.  We have also derived
the (well-behaved) vortex density distribution leading to $k$-scaling 
in the large-N limit.

   The essential feature of the center confinement mechanism, aside from
center dominance, is the finite correlation length of center flux in
a plane. Independent fluctuations of the center flux in finite regions 
guarantees an area law to Wilson loops of non-zero N-ality, with an asymptotic
string tension that depends \emph{only} on the N-ality. 
The necessary randomizing of center flux at finite distances
can be achieved via vortex percolation,
as in $Z_2$ or Yang-Mills lattice gauge theory, or from the disordering
due to a center monopole plasma, as in compact $Z_N$ or SU(N) gauge
theory at large $N$.  At moderate values of $N$, perhaps as low as $N=3$,
it is likely that both effects play a role.

\acknowledgments{ 

  We thank Dmitri Diakonov and Herbert Neuberger for helpful
discussions.  J.G.\ is grateful for the  hospitality provided by
the Niels Bohr Institute, where his work on this project was
carried out.  Our research is supported 
in part by the U.S. Department of 
Energy under Grant No.\ DE-FG03-92ER40711 (J.G.), and the Slovak Grant 
Agency for Science, Grant No. 2/7119/2000 (\v{S}.O.).
Our collaborative effort was also supported by NATO Collaborative Linkage
Grant No.\ PST.CLG.976987. }

\end{document}